\documentclass[pra,twocolumn,showpacs,nopreprintnumbers,amsmath,amssymb]{revtex4}
\usepackage{graphicx}
\usepackage{dcolumn}
\usepackage{bm}

\newtheorem{df}{~~~{\bf Definition}}[section]
\newtheorem{thm}[df]{~~~{\bf Theorem}}  
\newtheorem{pp}[df]{~~~{\bf Proposition}}  
\newtheorem{lm}[df]{~~~{\bf Lemma}}

\newcommand{\qed}{\raisebox{.6ex}{\fbox{\rule[0.0mm]{0mm}{0.8mm}}}}

\newcommand{\ket}[1]{%
           | #1 \rangle}

\newcommand{\bra}[1]{%
           \langle #1 |}

\newcommand{\braket}[2]{%
           \langle #1 | #2 \rangle}

\newcommand{\ketbra}[2]{%
           | #1 \rangle \langle #2 |}

\begin{document}

\preprint{}

\title{
Bound-state eigenenergy 
outside and inside the continuum \\
for unstable multilevel systems
}

\author{Manabu Miyamoto}%
 \email{miyamo@hep.phys.waseda.ac.jp }
\affiliation{%
Department of Physics, Waseda University, 3-4-1 Okubo, Shinjuku-ku, 
Tokyo 169-8555,  Japan 
}%

\date{\today}

\begin{abstract}
The eigenvalue problem for the dressed bound-state of unstable multilevel systems 
is examined both outside and inside the continuum, based on the $N$-level Friedrichs model 
which describes the couplings between the discrete levels and the continuous spectrum. 
It is shown that 
a bound-state eigenenergy always exists 
below each of the discrete levels that lie outside the continuum. 
Furthermore, by strengthening the couplings gradually, 
the eigenenergy corresponding to each of the discrete levels inside the continuum 
finally emerges. 
On the other hand, the absence of the eigenenergy inside the continuum 
is proved in weak but finite coupling regimes, provided that 
each of the form factors that determine 
the transition between some definite level and the continuum 
does not vanish at that energy level. 
An application to the spontaneous emission process for the hydrogen atom 
interacting with the electromagnetic field is demonstrated. 
\end{abstract}

\pacs{42.50.Vk, 42.50.Md}  

\maketitle

\section{Introduction} \label{sec:0}

The theoretical description of the unstable quantum system often refers to 
the system of a finite level coupled with the spectral continuum. 
In weak coupling regimes, 
the initial state localized at the finite level undergoes exponential decay 
\cite{Nakazato(1996)}. 
However, by changing the couplings to stronger regimes, instead of the total decay 
a partial one can occur \cite{Gaveau(1995)}. 
This means that 
the superposition between the states localized at the finite level and the continuum 
forms the dressed bound state, that is, a bound eigenstate extended over the total Hilbert space. 
The formation of the bound eigenstate 
is of great interest in the study of the various systems having to do with such matters as 
the photodetachment of electrons from negative ions 
\cite{Rzazewski(1982),Nakazato(2003)}
and the spontaneous emission of photons from atoms in photonic crystals 
\cite{John(1990),Kofman(1994),Wang(2003)}. 
It is then clarified that 
the energy of the bound eigenstate depends not only on 
the strength of the couplings but also on 
the relative location between the electron bound-energy and 
the detachment threshold \cite{Rzazewski(1982),Nakazato(2003)}, 
or 
between the energy of the atomic frequency  
and the continuum edge of the radiation frequency \cite{Kofman(1994),Wang(2003)}. 
Further research has been directed to those eigenstates, 
aiming at the decoherence control \cite{Antoniou(2004),Pellegrin(2005)}

In these analyses, however, single-level systems are treated often, 
while multilevel systems are less examined. 
In the latter, some peculiar time evolutions are theoretically observed: 
steplike decay \cite{Frishman(2001)}, decaying oscillation \cite{Antoniou(2003)}, 
and various long-time nonexponential decays \cite{Miyamoto(2004),Miyamoto(2005)}. 
These peculiarities are never found in single-level approaches. 
Furthermore, 
to the author's knowledge, 
the possibility of a bound-state eigenenergy ``inside'' the continuum 
has not been studied 
except in a special multilevel case where 
all form factors are assumed to be identical 
\cite{Antoniou(2004),Davies(1974)}.

In the present paper, we attempt to examine 
the eigenvalue problem for the dressed bound-state in multilevel cases, 
based on the $N$-level Friedrichs model \cite{Friedrichs(1948),Exner(1985)}, 
allowing some class of form factors, including identical cases. 
We show that for the discrete energy levels lying outside the continuum, 
the bound-state eigenenergy always remains below each of them. 
Moreover, by increasing the couplings, 
the bound-state eigenenergy corresponding to each of the discrete levels inside the continuum 
can emerge out of that continuum. 
For the bound-state eigenenergy inside the continuum, 
we can only 
prove its absence in weak coupling cases 
under the condition that the form factors do not vanish at the energy of each level. 
This result is just an extension of the lemma already proved for a system with 
identical form factors \cite{Davies(1974)}. 
An upper bound of the coupling constant for the case of no bound-state eigenenergy being 
inside the continuum is also obtained explicitly. 
We apply this result to the spontaneous emission process for the hydrogen atom 
under the four-level approximation.

In the next section, 
we introduce the $N$-level Friedrichs model and its eigenvalue problem. 
In Sec. \ref{sec:3}, 
we consider the eigenvalues outside the continuum, 
with resort to the perturbation theory about the eigenvalue of hermitian matrix. 
The discussion developed here helps us to undertake 
the problem for the inside case, 
which is argued in Sec. \ref{sec:4}. 
Concluding remarks are given in Sec. \ref{sec:5}. 
We also present an appendix where both the small and large energy behaviors of the energy shift 
are studied in detail.

\section{The $N$-level Friedrichs model and the eigenvalue problem} 
\label{sec:2}

The $N$-level Friedrichs model describes the $N$-level system coupled with the continuum system. 
The total Hamiltonian $H$ is defined by 
\begin{equation}
H=H_0 + \lambda V, 
\label{eqn:2.60}
\end{equation}
where $\lambda \in \mathbb{R}$ is the coupling constant. 
We here define the free Hamiltonian $H_0$ as 
\begin{equation}
H_0 =\sum_{n=1}^N \omega_n \ketbra{n}{n} 
+\int_\Omega \omega \ketbra{\omega}{\omega} \rho(\omega) d\omega ,
\label{eqn:2.40}
\end{equation}
where it was assumed that $\omega_1 \leq \omega_2 \leq \ldots \leq \omega_N$. 
$\ket{n}$ and $\ket{\omega}$ satisfy the orthonormality condition: 
$\langle n | n^{\prime} \rangle = \delta_{n n^{\prime}}$, 
$\braket{ \omega }{ \omega^{\prime} } =\delta (\omega - \omega^{\prime})/\rho (\omega )$, 
and $\langle n | \omega \rangle = 0 $, 
where $\delta_{n n^{\prime}}$ is Kronecker's delta and 
$\delta (\omega - \omega^{\prime})$ is Dirac's delta function. 
$\rho (\omega )$ is a nonnegative function interpreted as, 
e.g., an electromagnetic density of mode, 
and $\Omega=\{\omega | \rho (\omega)\neq 0 \}$ is 
a specific region, like the energy band allowed by the electromagnetic mode. 
The interaction Hamiltonian $V$ describing the couplings between $\ket{n}$ and $\ket{\omega}$ is 
\begin{equation}
V = 
\sum_{n=1}^{N}
\int_\Omega
\left[ v_n (\omega ) | \omega \rangle \langle n | +
v_n^* (\omega ) | n \rangle \langle \omega | \right] \rho(\omega) d \omega ,
\label{eqn:2.70}
\end{equation}
where ($^*$) denotes the complex conjugate 
and $v_n (\omega )$ is the form factor characterizing 
the transition between $| n \rangle$ and $| \omega \rangle$. 
We here assumed that $v_n \in L^2 (0, \infty )$, i.e., 
\begin{equation}
\int_\Omega |v_n (\omega )|^2 \rho(\omega) d\omega <\infty. 
\label{eqn:2.80}
\end{equation}
For clarity of discussion below, we assume that $\rho (\omega )=1$ for $\omega \geq 0$
and $0$ otherwise, so that $\Omega =[0, \infty)$. 
Then, we merely write $\int_\Omega$ by $\int_{0}^{\infty}$, and 
the outside of the continuum means the half line $(-\infty, 0)$. 
An extension of $\Omega$ to more general cases, such as gap structures, 
is not difficult, however, our facilitation could extract the essential of the matter.

Let us next set up the eigenvalue problem for this model. 
We suppose that 
the eigenstate corresponding to the eigenvalue $E$ 
is of the form $\ket{u_E} =\sum_{n=1}^N c_n \ket{n} 
+\int_0^{\infty} f(\omega) \ket{\omega} d\omega $, and it is normalizable, i.e., 
\cite{pointspectrum}
\begin{equation}
\braket{u_E}{u_E} = \sum_{n=1}^{N}  |c_n |^2 
+ \int_0^{\infty} |f(\omega) |^2 d\omega < \infty .
\label{eqn:2.25}
\end{equation}
Then, the eigenequation $H\ket{u_E} =E \ket{u_E}$ is equivalent to the following ones,
\begin{equation}
\omega_n c_n + \lambda \int_0^{\infty} v_n^* (\omega) f(\omega) d\omega  =Ec_n , 
~~~
\forall n =1, \ldots, N, 
\label{eqn:3.10}
\end{equation}
\begin{equation}
\omega f(\omega ) + \lambda \sum_{n=1}^{N} c_n v_n (\omega ) =E f(\omega ) .
\label{eqn:3.20}
\end{equation}
Equation (\ref{eqn:3.20}) immediately implies 
\begin{equation}
f(\omega)=- \lambda \frac{\sum_{n=1}^{N} c_n v_n (\omega )}{\omega -E}. 
\label{eqn:3.25}
\end{equation}
By setting this into Eq. (\ref{eqn:2.25}), we have the normalization condition 
\begin{equation}
\int_{0}^{\infty } 
|f(\omega)|^2 d\omega 
=\lambda^2 \int_{0}^{\infty } 
\frac{|\sum_{n=1}^{N} c_n v_n (\omega ) |^2}{|\omega -E|^2} d\omega 
< \infty ,
\label{eqn:3.30}
\end{equation}
which is the essential of the localization of dressed bound-state.


\section{Bound-state eigenenergy outside the continuum} 
\label{sec:3}

We first review the results on the negative-eigenvalue problem for $N=1$, the single-level case 
\cite{Horwitz(1971)}. If $E<0$, the integral in Eq. (\ref{eqn:3.30}) always converges 
under the condition (\ref{eqn:2.80}). In fact,
\begin{equation}
|c_1|^2
\int_{0}^{\infty } 
\frac{|v_1 (\omega ) |^2}{|\omega +|E| |^2} d\omega 
\leq 
\frac{|c_1|^2}{|E|^2}
\int_{0}^{\infty } |v_1 (\omega ) |^2 d\omega < \infty.
\label{eqn:3.33}
\end{equation}
Thus, the substitution of $f(\omega)$ into Eq. (\ref{eqn:3.10}) is allowed. 
By introducing the function $\kappa (E)$ as 
\begin{equation}
\kappa(E)=\omega_1 - \lambda^2 
\int_{0}^{\infty } 
\frac{|v_1 (\omega ) |^2}{\omega -E} d\omega,
\label{eqn:3.34}
\end{equation}
Eq. (\ref{eqn:3.10}) reads \cite{c_1}
\begin{equation}
\kappa (E)=E,
\label{eqn:3.35}
\end{equation}
which is either an algebraic or transcendental equation of $E$, depending on $v_1 (\omega)$. 
$\kappa(E)$ has two important properties as follows, 
\begin{equation}
\kappa(E') \geq \kappa(E), ~~~ \mbox{and} ~~~ \kappa(E)\leq \omega_1, 
\label{eqn:3.36}
\end{equation}
for all $E$ and $E'$ satisfying $E' \leq E<0$. 
The former means that $\kappa(E)$ is monotone decreasing in $E$. 
Therefore,  
there is only one solution (negative eigenvalue) $E$ of Eq. (\ref{eqn:3.35}) if and only if 
\begin{equation}
\lim_{E\uparrow 0} \kappa(E)=
\omega_1 - \lim_{E\uparrow 0} \lambda^2 
\int_{0}^{\infty } 
\frac{|v_1 (\omega ) |^2}{\omega -E} d\omega < 0. 
\label{eqn:3.37}
\end{equation}
When $E>0$, $E$ should be a zero of $v_1 (\omega)$ 
so that Eq. (\ref{eqn:3.30}) holds. 
This is discussed in detail in Sec. \ref{sec:4}.

Let us now turn to the $N$-level case. 
Corresponding to Eq. (\ref{eqn:3.33}), this time we have that
\begin{equation}
\int_{0}^{\infty } 
\frac{|\sum_{n=1}^{N} c_n v_n (\omega ) |^2}{|\omega +|E| |^2} d\omega 
\leq 
\frac{\sum_{n=1}^{N} \int_{0}^{\infty} | v_n (\omega)|^2 d\omega
}{|E|^2 }
< \infty ,
\label{eqn:3.1.10}
\end{equation}
and Eq. (\ref{eqn:3.30}) is satisfied again, where we used that $\sum_{n=1}^{N} |c_n |^2 \leq1$. 
Substituting Eq. (\ref{eqn:3.25}) into (\ref{eqn:3.10}), 
one obtains 
\begin{equation}
\sum_{n^{\prime }=1}^{N} 
\bigl[ 
\omega_n \delta_{nn^{\prime }} -\lambda^2 S_{nn^{\prime }}(E) 
\bigr]
c_{n^{\prime }} =E c_n,
\label{eqn:3.1.20}
\end{equation}
where
\begin{equation}
S_{nn^{\prime }}(z) 
=\int_{0}^{\infty } 
\frac{v_n^* (\omega ) v_{n^{\prime}} (\omega )}{\omega -z} d\omega ,
\label{eqn:3.1.30}
\end{equation}
with $z \in \mathbb{C}\backslash [0,\infty)$. 
For a later convenience, 
we here introduce an $N \times N$ matrix $S(z)$ with the components 
$S_{nn^{\prime }}(z) $. 
Note that $S(E)$ for $E<0$ turns out to be a Gram matrix \cite{MatrixAnalysis}, 
which is positive semidefinite. 
One obtains the following property of $S(E)$:

\begin{lm}\label{pp:4.1}
$S(E^{\prime}) \leq S(E)$ for $E^{\prime} \leq E <0$. 
\end{lm}

{\sl Proof} :  
We have that
\begin{equation}
S_{nn^{\prime}}(E) - S_{nn^{\prime}}(E^{\prime}) 
=
(E -E^{\prime} )
T_{nn^{\prime}}(E,E^{\prime}) , 
\label{eqn:3.1.60}
\end{equation}
for all $E$ and $E^{\prime}$ satisfying $E^{\prime} \leq E <0$. 
We here introduce the matrix $T(E ,E^{\prime})$ whose components are
\begin{equation}
T_{nn^{\prime}}(E,E^{\prime})
:=
\int_{0}^{\infty } 
\frac{v_n^* (\omega ) v_{n^{\prime}} (\omega )}
{(\omega -E)(\omega -E^{\prime})} d\omega .
\label{eqn:3.1.70}
\end{equation}
Note that since $T(E ,E^{\prime})$ is a Gram matrix, 
it is positive semidefinite. Therefore the proof is completed. 
\qed

We also introduce the matrices $K_0$ and $K(E)=K_0-\lambda^2S(E)$ with components 
\begin{equation}
{K_0}_{nn^{\prime}} :=
\omega_n \delta_{nn^{\prime }} ,
\label{eqn:3.1.35}
\end{equation}
and\begin{equation}
K_{nn^{\prime}}(E) :=
\omega_n \delta_{nn^{\prime }} -\lambda^2 S_{nn^{\prime }}(E) ,
\label{eqn:3.1.40}
\end{equation}
respectively. For any $E<0$, $K(E)$ becomes a hermitian matrix, and thus 
there are $N$ eigenvalues of $K(E)$. We denote them by $\{\kappa_n (E) \}_{n=1}^{N}$, 
where $\kappa_1 (E) \leq \kappa_2 (E) \leq \ldots \leq \kappa_N (E)$. 
The existence of a nontrivial solution $\{ c_n \}$ of Eq. (\ref{eqn:3.1.20}) 
is guaranteed if and only if there exists a negative $E$ to satisfy 
\begin{equation}
\kappa_n (E) =E ,
\label{eqn:3.1.50}
\end{equation}
for a certain integer $n$. 
As in the former part of Eq. (\ref{eqn:3.36}), 
$\kappa_n(E)$ has the following property:

\begin{lm}\label{pp:4.2}
For any fixed $n$, 
$\kappa_n (E^{\prime}) \geq \kappa_n (E)$ for $E^{\prime} \leq E <0$. 
\end{lm}

{\sl Proof} :  
We see from Eq. (\ref{eqn:3.1.60}) that 
\begin{equation}
K(E) - K(E^{\prime}) 
%
%
=
-(E -E^{\prime} )\lambda^2 T(E ,E^{\prime}) \leq 0,
\label{eqn:3.1.80}
\end{equation}
for $E^{\prime} \leq E <0$. 
Then, by using the Theorem 4.3.1 in Ref. \cite{MatrixAnalysis}, 
the following inequality between the eigenvalues of $K(E)$, $K(E')$, and $T(E ,E^{\prime})$
holds \cite{inequality}, 
\begin{eqnarray}
&&\kappa_n (E^{\prime}) -(E -E^{\prime} )\lambda^2 \tau_N (E ,E^{\prime})
\nonumber \\
&&\leq
\kappa_n (E)
\leq
\kappa_n (E^{\prime}) -(E -E^{\prime} )\lambda^2 \tau_1 (E ,E^{\prime}),
\label{eqn:3.1.100}
\end{eqnarray}
where $\tau_n (E ,E^{\prime})$ denotes the $n$-th eigenvalue of $T(E ,E^{\prime})$. 
Note that since $T(E ,E^{\prime})\geq 0$, all $\tau_n (E ,E^{\prime})\geq0$. 
Then, $-(E -E^{\prime} )\tau_1 (E ,E^{\prime}) \leq 0$ for $E \geq E^{\prime} $, 
and the inequality 
\begin{equation}
\kappa_n (E) \leq \kappa_n (E^{\prime}),
\label{eqn:3.3.110}
\end{equation}
immediately follows from the last part of Eq. (\ref{eqn:3.1.100}). 
\qed

We also have the statement below, which corresponds to the latter part of Eq. (\ref{eqn:3.36}). 

\begin{lm}\label{pp:4.3}
For any fixed $n$, 
$ \kappa_n (E) \leq \omega_n$ for all $E<0$, and 
$\displaystyle \lim_{E \to -\infty} \kappa_n (E) =\omega_n$. 
\end{lm}

{\sl Proof} :  
From Eq. (\ref{eqn:3.1.40}) and Theorem 4.3.1 in Ref. \cite{MatrixAnalysis} 
again, one obtains that 
\begin{equation}
\omega_n - \lambda^2 \sigma_N (E)
\leq
\kappa_n (E)
\leq
\omega_n - \lambda^2 \sigma_1 (E) ,
\label{eqn:3.1.140}
\end{equation}
where $\sigma_n (E)$ denotes the $n$-th eigenvalue of $S(E)$.  
If we recall the fact that $S(E)\geq 0$ implies $\sigma_n (E) \geq 0$ 
for every $n$, the above inequality reads
\begin{equation}
0\leq \lambda^2 \sigma_1 (E) 
\leq 
\omega_n - \kappa_n (E)
\leq 
\lambda^2 \sigma_N (E) .
\label{eqn:3.1.150}
\end{equation}
Asymptotic behavior of the right-hand side of the above can be evaluated from 
Eq. (\ref{eqn:3.1.30}) as 
\begin{equation}
\sigma_N (E) \leq \mathrm{tr} (S(E)) 
\leq 
\frac{1}{|E| } \sum_{n=1}^{N} \int_{0}^{\infty} | v_n (\omega)|^2 d\omega
\rightarrow 0 ,
\label{eqn:3.1.160}
\end{equation}
as $E \rightarrow -\infty$, and thus the lemma is proved. 
\qed


Therefore, summarizing Lemmas \ref{pp:4.2} and \ref{pp:4.3}, 
we obtain 

\begin{thm}\label{thm:4.1}
If $\lim_{E\uparrow 0} \kappa_n (E) <0$ up to $n=M$, 
then each of the $\kappa_n (E)$ for $n=1, \ldots, M$ 
intersects $E$ only once, so that $M$ negative eigenenergies of $H$ exist. 
In particular, if $H_0$ has $N_-$ negative eigenenergies, i.e., 
$\omega_n  <0$ up to $n=N_-$,  then $N_-$ negative eigenenergies of $H$, 
denoted by $E_n$, exist and satisfy $E_n \leq \omega_n$. 
\end{thm}


We also see from Eq. (\ref{eqn:3.1.140}) that 
\begin{equation}
\kappa_n (E)
\leq 
\omega_n - \lambda^2 \sigma_1 (E). 
\label{eqn:3.1.170}
\end{equation}
This means that 
when $|\lambda |$ is large enough, every $\kappa_n (E)$, 
even originating from a positive $\omega_n$, becomes negative, 
unless $\sigma_1 (E)=0$, i.e., the $v_n (\omega)$'s are linearly dependent \cite{MatrixAnalysis}. 
More precisely, the following statement holds.

\begin{pp}\label{pp:4.4}
Suppose that only $N_{\rm ind}$ form factors are linearly independent 
among them. 
Then, it follows that for any $E<0$, 
\begin{equation}
-\lambda^2 \sigma_{N+1-n} (E) +\omega_1
\leq
\kappa_n (E)
\leq
-\lambda^2 \sigma_{N+1-n} (E) +\omega_N , 
\label{eqn:3.1.180}
\end{equation}
and $\sigma_{N+1-n} (E)\neq 0$ for $n=1, \ldots, N_{\rm ind}$, while
\begin{equation}
\omega_1
\leq
\kappa_n (E)
\leq
\omega_N , 
\label{eqn:3.1.190}
\end{equation}
for $n=N_{\rm ind} +1, \ldots, N$. 
Therefore, only the first $N_{\rm ind}$ eigenvalues of $K(E)$ are ensured 
to be negative as $|\lambda|$ goes to infinity without regard to 
the location of $\{ \omega_n \}_{n=1}^N$. 
\end{pp}

{\sl Proof} :  
Taking $-\lambda^2 S(E)$ as the unperturbed part of $K(E)$, we obtain Eq. (\ref{eqn:3.1.180}) for all $n$. 
Note that if only $N_{\rm ind}$ form factors are linearly independent, 
it holds that $\sigma_{m} (E)=0$ for $m=1, \ldots, N-N_{\rm ind}$ 
and otherwise does not vanish. 
Then, the assertion is proved straightforwardly. 
\qed

\begin{figure}
\hspace*{-11mm}
\rotatebox{270}{
\includegraphics[width=0.33\textwidth]{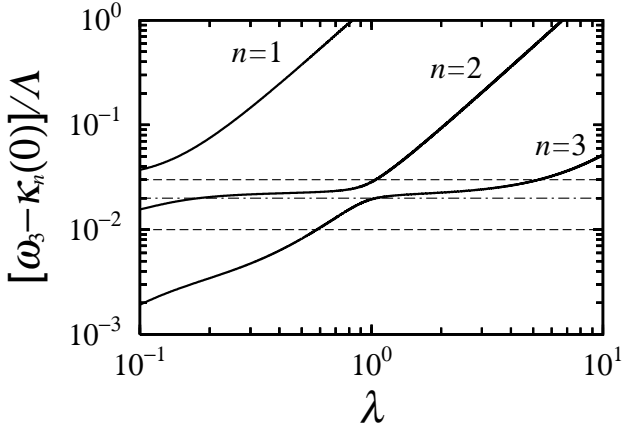}
}
\caption{\label{fig:figure1}
$\omega_3-\kappa_n(0)$ for $n=1$, $2$, $3$ 
(three solid lines) for a three-level system with form factors (\ref{eqn:3.1.200}), 
plotted against $\lambda$, and $\omega_3-\omega_1$, $\omega_3-\omega_2$ 
(two dashed lines), and $\omega_3$ (dot-dashed line), for reference, 
where $\omega_3-\omega_1 >\omega_3 > \omega_3-\omega_2$. 
Three different regions are distinguished, corresponding to 
the number of solid lines satisfying $\omega_3-\kappa_n(0)>\omega_3$, 
that is, just the number of negative eigenenergies of $H$, by Theorem \ref{thm:4.1}.
%
%
}
\end{figure}
\begin{figure}
\hspace*{-11mm}
\rotatebox{270}{
\includegraphics[width=0.33\textwidth]{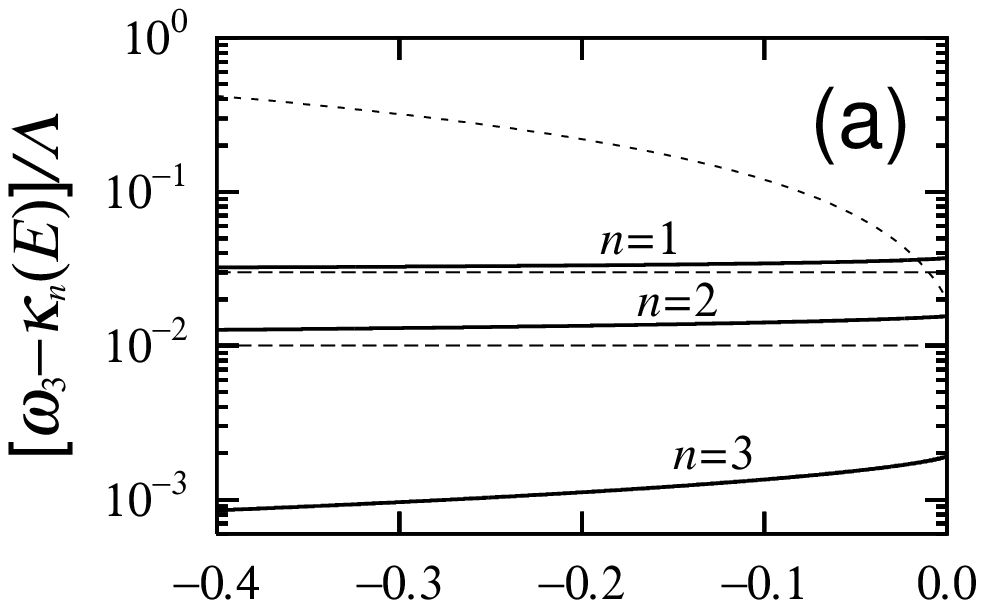}
}

\vspace*{-10mm}
\hspace*{-11mm}
\rotatebox{270}{
\includegraphics[width=0.33\textwidth]{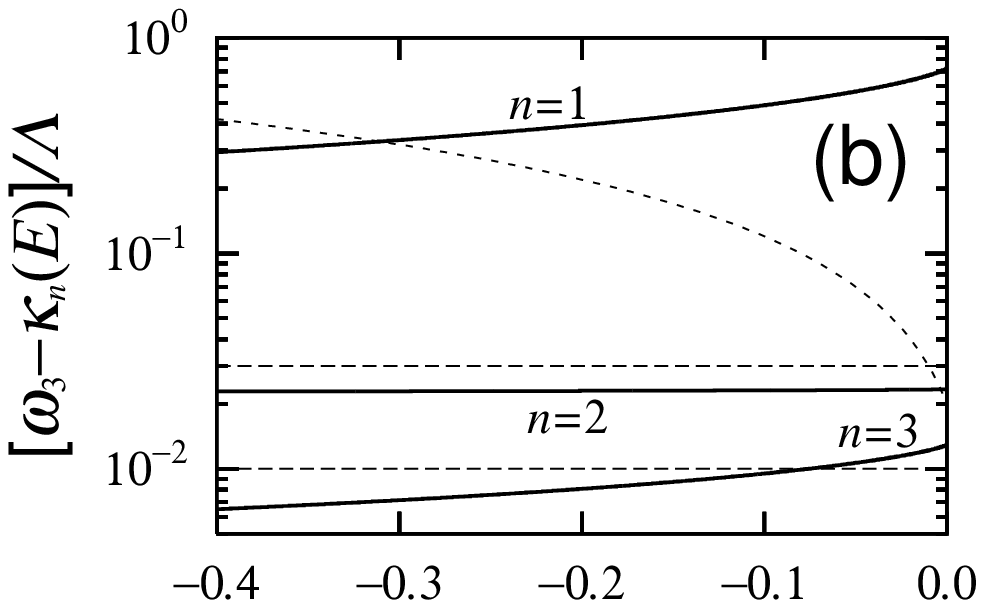}
}

\vspace*{-10mm}
\hspace*{-11mm}
\rotatebox{270}{
\includegraphics[width=0.33\textwidth]{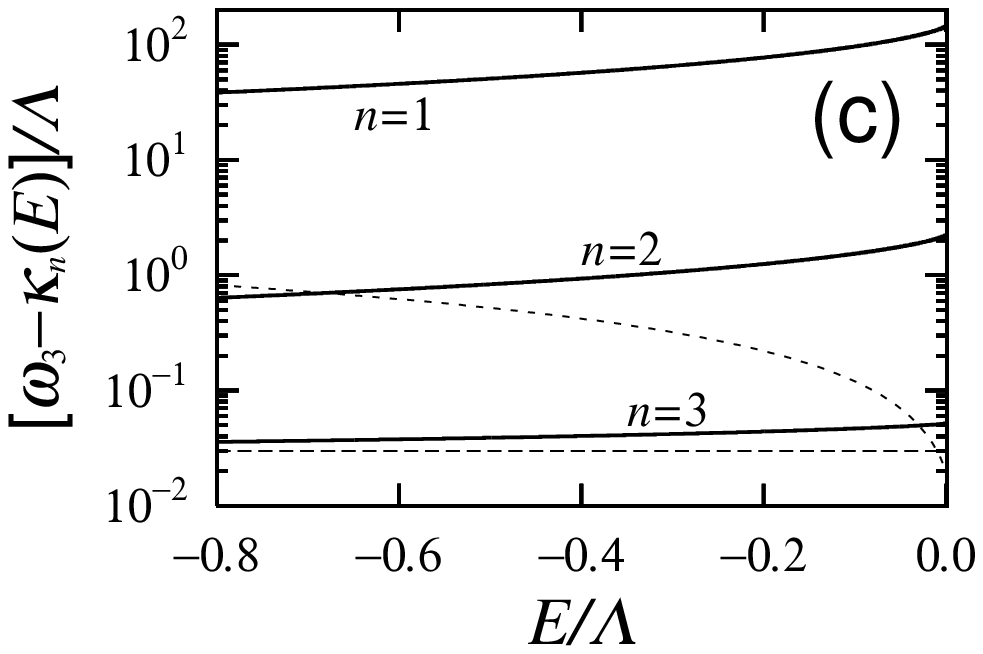}
}
\caption{\label{fig:figure2} 
$\omega_3-\kappa_n(E)$ for $n=1$, $2$, $3$ 
(three solid lines) for a three-level system of the form factors (\ref{eqn:3.1.200}), 
$\omega_3-E$ (short-dashed line), and $\omega_3-\omega_1$ and $\omega_3-\omega_2$  
(two dashed lines). 
We plot them in $\lambda=0.1$, $0.7$, and $10.0$, in (a), (b), and (c), respectively, 
choosing the parameters as 
$\omega_1/\Lambda =-0.01$, $\omega_2/\Lambda=0.01$, and $\omega_3/\Lambda=0.02$. 
(a) 
For a relatively small $\lambda$, 
only $\omega_3-\kappa_1(E)$ intersects $\omega_3-E$ 
at $E/\Lambda \simeq-0.02$, 
which is predicted by Theorem \ref{thm:4.1}. 
Thus there is one negative eigenenergy. 
One also sees that $\omega_3-\kappa_1(E)$ and $\omega_3-\kappa_2(E)$ still lie closely above 
the asymptotes $\omega_3-\omega_1$ and $\omega_3-\omega_2$, respectively. 
(b) 
Both $\omega_3-\kappa_1(E)$ and $\omega_3-\kappa_2(E)$ intersect $\omega_3-E$ 
in the vicinity of $E=-0.3$ and $0.0$, respectively, 
so that there are two negative eigenenergies. 
(c) 
All $\omega_3-\kappa_n(E)$ for $n=1$, $2$, and $3$ intersect $\omega_3-E$, 
and thus three negative eigenenergies exist. 
In this figure, only two intersections for $n=2$ and $3$ are depicted. 
}

\end{figure}

To illustrate the emergence of the negative eigenenergies, 
described in Theorem \ref{thm:4.1} and Proposition \ref{pp:4.4}, 
let us consider the three-level system especially in the case where 
$\omega_1 <0$ while $\omega_2 >0$ and $\omega_3>0$. 
We also choose three form factors, such as
\begin{equation}
v_n (\omega)=\Lambda^{1/2} \frac{\sqrt{\omega/\Lambda} [1+a_n (\omega/\Lambda)^{2(n-1)}]}
{[1+(\omega/\Lambda)^2]^{1+n}} ,
\label{eqn:3.1.200}
\end{equation}
where $\Lambda$ is the cut off constant, and $a_n$ is a parameter. 
The form factors described by such algebraic functions are often found in various systems 
involving the process of the spontaneous emission of photons from the hydrogen atom 
\cite{Seke(1994),Facchi(1998)}, 
the photodetachment of electrons from negative ions 
\cite{Rzazewski(1982),Nakazato(2003),Haan(1984)}, 
and quantum dots \cite{Antoniou(2001)}. 
In the calculation depicted in Figs. \ref{fig:figure1} and \ref{fig:figure2}, 
we have chosen a set of parameters 
$\omega_1 /\Lambda =-0.01$, $\omega_2 /\Lambda =0.01$, and $\omega_3 /\Lambda =0.02$, 
and $a_1=0.0$, $a_2=2.0$, $a_3=1.0$. 
These choices for $a_n$ guarantee linear independency among $v_n$'s , 
so that $N_{\rm ind}=3$.

Figure \ref{fig:figure1}
shows $\omega_3-\kappa_n(0)$ for $n=1,2,3$, changing $\lambda$ from $0.1$ to $10.0$, 
and $\omega_3 -\omega_1$, $\omega_3 -\omega_2$, (two dashed lines) 
and $\omega_3$ (dot-dashed line) for reference. 
The latter satisfy the relation that $\omega_3 -\omega_1>\omega_3>\omega_3 -\omega_2>0$. 
One may recognize three different regions in this figure: 
for small $\lambda \lesssim 0.2$, 
one inequality $\omega_3-\kappa_1 (0)>\omega_3$, i.e., $\kappa_1 (0)<0$, holds. 
In the next region $0.2 \lesssim \lambda \lesssim 1.0$, 
two inequalities, $\omega_3-\kappa_1 (0)>\omega_3$ and $\omega_3-\kappa_2 (0)>\omega_3$, hold. 
For $\lambda \gtrsim 1.0$ the last region, three inequalities, 
$\omega_3-\kappa_n (0)>\omega_3$ for all $n=1, 2, 3$, are satisfied. 
Therefore, according to Theorem \ref{thm:4.1}, 
one sees that one, two, and three negative eigenenergies of $H$ exist 
in the first, second, and third regions, respectively. 
It is worth noting that the appearance of the negative eigenenergy 
in the first region merely occurs from 
the fact that $\omega_1<0$ (see, the latter part of Theorem \ref{thm:4.1}), 
whereas that in other regions could be understood as 
a strong-coupling effect (Proposition \ref{pp:4.4}).

Figure \ref{fig:figure2}
shows three curves of $\omega_3-\kappa_n(E)$ for $n=1, 2, 3$ (three solid lines) 
and $\omega_3-E$ (short dashed line), plotted against $E$. 
An intersection of the former and the latter means an emergence of a negative eigenenergy. 
We also plot the asymptotes $\omega_3 -\omega_1$ and $\omega_3 -\omega_2$ (two dashed lines), 
to which $\omega_3-\kappa_1(E)$ and $\omega_3-\kappa_2(E)$ are close from above 
as $E\to -\infty$, respectively (see, Lemma \ref{pp:4.3}). 
Figures \ref{fig:figure2} (a), \ref{fig:figure2} (b), and \ref{fig:figure2} (c), 
are in the cases where 
$\lambda =0.1$, which belongs to the first region, $\lambda=0.7$, of the second one, 
and $\lambda=10.0$, of the last one, respectively. See Fig. \ref{fig:figure1}. 
It is seen 
in Fig.\ \ref{fig:figure2} (a) that 
$\omega_3-E$ intersects $\omega_3-\kappa_1(E)$ only, 
so that there is one negative eigenenergy. 
In Fig. \ref{fig:figure2} (b), one distinguishes the two intersections between 
$\omega_3-E$ and $\omega_3-\kappa_1(E)$, 
and between $\omega_3-E$ and $\omega_3-\kappa_2(E)$. Thus two negative eigenenergies appear. 
The intersection between the latter pair still lies around $E=0.0$. 
In Fig.\ \ref{fig:figure2} (c), where a relatively large $\lambda$ was chosen, 
$\omega_3-E$ finally intersects all three lines, $\omega_3-\kappa_n(E)$ for $n=1, 2, 3$, 
which tells us three negative eigenenergies exist.


\section{Absence of bound-state eigenenergy inside the continuum} 
\label{sec:4}

Let us next examine the nonnegative-eigenvalue problem 
for Eqs. (\ref{eqn:3.10}) and (\ref{eqn:3.20}). 
In this case, the normalization condition (\ref{eqn:3.30}) does not hold automatically, 
unlike the case where $E<0$, 
because of a possible divergence of $f(\omega)$ at $\omega=E$. 
Before going to the $N$-level case, let us first observe the single-level one. 
Except in the trivial case where $c_1 =0$, 
the condition (\ref{eqn:3.30}) for an eigenvalue $E \geq0$, if any, 
imposes the nontrivial condition or constraint that
\begin{equation}
v_1 (E)=0,
\label{eqn:3.2.55a}
\end{equation}
where we assume some extent of the smoothness of $v_1 (\omega )$ \cite{zeros}. 
Then, $f(\omega)=-\lambda c_1 v_1 (\omega )/(\omega-E)$ 
is ensured to be square integrable, and Eq. (\ref{eqn:3.10}) reads 
\begin{equation}
\omega_1 -\lambda^2 \int_0^{\infty }\frac{|v_1 (\omega)|^2}{\omega -E} d\omega =E.
\label{eqn:3.2.55b}
\end{equation}
To find the solution $E$ of Eq. (\ref{eqn:3.2.55b}), one may attempt to interpret it 
as an intersection between the left-hand and the right-hand sides, 
as in Eq. (\ref{eqn:3.35}). However, this approach seems impossible at first, 
because the left-hand side of Eq. (\ref{eqn:3.2.55b}) is not well defined for a general $E$ 
except such points satisfying Eq. (\ref{eqn:3.2.55a}). 
This matter can be solved by alternatively considering the following equation, 
\begin{equation}
\omega_1 -\lambda^2 P\int_0^{\infty }\frac{|v_1 (\omega)|^2}{\omega -E} d\omega =E,
\label{eqn:3.2.55c}
\end{equation}
that is obtained 
from Eq. (\ref{eqn:3.2.55b}) 
by replacing $\int_0^{\infty }\frac{|v_1 (\omega)|^2}{\omega -E} d\omega$ 
with its principal value $P\int_0^{\infty }\frac{|v_1 (\omega)|^2}{\omega -E} d\omega$. 
In this case, the left-hand side can make sense for a general $E$, 
and we can treat $E$ as an independent variable. 
If we find the solution $E$ of Eq. (\ref{eqn:3.2.55c}), 
and furthermore if it satisfies Eq. (\ref{eqn:3.2.55a}), 
it becomes a true solution of the original equation (\ref{eqn:3.2.55b}). 
Indeed, in such a situation, we have that 
$\int_0^{\infty }\frac{|v_1 (\omega)|^2}{\omega -E} d\omega
=P\int_0^{\infty }\frac{|v_1 (\omega)|^2}{\omega -E} d\omega$, and thus 
Eq. (\ref{eqn:3.2.55c}) just reproduces Eq. (\ref{eqn:3.2.55b}).

In the $N$-level cases, the condition (\ref{eqn:3.30}) for an eigenvalue $E \geq0$, if any, 
can be translated into the equivalent condition 
for both the coefficients $\{ c_n \}_{n=1}^N$ and $E$, that is,
\begin{equation}
\sum_{n=1}^{N} c_n v_n (E) 
=0.
\label{eqn:3.2.10}
\end{equation}
Under this condition, we can safely substitute Eq. (\ref{eqn:3.25}) into Eq. (\ref{eqn:3.10}). 
However, similarly to Eq. (\ref{eqn:3.2.55c}), we consider the alternative equation 
in the $N$-level cases as 
\begin{equation}
\sum_{n^{\prime} =1}^{N} [\omega_n \delta_{nn^\prime} -\lambda^2 D_{nn^\prime}(E)] 
c_{n^{\prime} }=Ec_n ,
\label{eqn:3.2.50}
\end{equation}
for $n=1, 2, \ldots, N$, where 
\begin{equation}
D_{n n^{\prime}} (E)
:=
P \int_0^{\infty }
\frac{v_n^* (\omega ) v_{n^{\prime}} (\omega)}{\omega -E}
d\omega,
\label{eqn:3.2.30}
\end{equation}
which are the components of the hermitian matrix $D(E)$ defined for all $E\geq 0$. 
One sees that Eq. (\ref{eqn:3.2.50}) has the same form as Eq. (\ref{eqn:3.1.20}), 
except the point that $S(E)$ ($E<0$) is replaced by $D(E)$ ($E\geq 0$). 
Then, we can implement a formulation in the matrix form, just as in the preceding section. 
In fact, 
the solutions of Eq. (\ref{eqn:3.2.50}) can be connected with those of Eq. (\ref{eqn:3.10}) 
under the condition (\ref{eqn:3.2.10}). 
We first note that 
\begin{eqnarray}
&&\hspace{-5mm}
P \int_0^{\infty }
\frac{v_n^* (\omega ) 
\sum_{{n^{\prime}}=1}^{N} c_{n^{\prime}} v_{n^{\prime}} (\omega)}{\omega -E}d\omega
\nonumber \\
&&=
\sum_{{n^{\prime}}=1}^{N} c_{n^{\prime}} 
P \int_0^{\infty }
\frac{v_n^* (\omega ) v_{n^{\prime}} (\omega)}{\omega -E}
d\omega,
\label{eqn:3.2.20b}
\end{eqnarray}
which is always valid for all $E$. 
Then, substituting this relation into Eq. (\ref{eqn:3.2.50}), we have
\begin{equation}
\omega_n c_n -\lambda^2 P \int_0^{\infty }
\frac{v_n^* (\omega ) 
\sum_{{n^{\prime}}=1}^{N} c_{n^{\prime}} v_{n^{\prime}} (\omega)}{\omega -E}d\omega=Ec_n,
\label{eqn:3.2.25}
\end{equation}
for $n=1, 2, \ldots, N$. 
For a comparison, see Eq. (\ref{eqn:3.10}) again. 
Therefore, if the solutions $E$ and $\{ c_n \}_{n=1}^N$ of Eq. (\ref{eqn:3.2.25}), 
i.e., Eq. (\ref{eqn:3.2.50}), 
satisfy the condition (\ref{eqn:3.2.10}), 
Eq. (\ref{eqn:3.2.25}) can reproduce Eq. (\ref{eqn:3.10}), so that 
the solutions of Eq. (\ref{eqn:3.2.25}) become the true ones of Eq. (\ref{eqn:3.10}).

Our procedure for finding 
the coefficients $\{ c_n \}_{n=1}^N$ and the nonnegative eigenvalue $E$ of $H$ 
that satisfy Eqs. (\ref{eqn:3.10}) and (\ref{eqn:3.20}) 
consists of the two steps: we first solve Eq. (\ref{eqn:3.2.50}), 
and then we check whether the solutions satisfy the condition (\ref{eqn:3.2.10}). 
For a later convenience, we introduce the hermitian matrix $K(E)$ for $E\geq 0$ 
whose components are defined by
\begin{equation}
K_{n n^{\prime}} (E)
:=
\omega_n \delta_{n n^{\prime}} -\lambda^2 D_{n n^{\prime}}(E).
\label{eqn:3.2.40}
\end{equation}
Then, the existence of a nontrivial solution of Eq. (\ref{eqn:3.2.50}) 
is ensured if and only if there exists a nonnegative $E$ to satisfy 
\begin{equation}
\kappa_n (E, \lambda) =E ,
\label{eqn:3.2.53}
\end{equation}
for a certain integer $n$, 
where $\{\kappa_n (E, \lambda) \}_{n=1}^{N}$ are the eigenvalues of $K(E)$, 
arranged in increasing order. 
To summarize again, if an eigenvalue of $K(E)$ is $E$, 
then it is an eigenvalue of $H$, provided that it also satisfies the condition (\ref{eqn:3.2.10}).

It is worth noting that the condition (\ref{eqn:3.2.10}) seems not 
necessarily to require the existence of a zero of $v_n (\omega)$, 
unlike the single-level case of Eq. (\ref{eqn:3.2.55a}). 
However, the following statement means that 
if $v_n (\omega_n )\neq 0$ for all $\omega_n >0$, the weak-coupling condition results in 
no positive eigenvalue of $H$ strictly.

\begin{thm}\label{thm:4.2}
Suppose that $H_0$ has $N_+$ positive eigenvalues without any degeneracy, 
and each $v_n (\omega )$ is an $L^2$-function 
of the form, $v_n (\omega )=\omega^{p_n} f_n (\omega )$, 
where $p_n >0$ and $f_n (\omega )$ is a $C^1$-function in $[0, \infty)$. 
Furthermore, it is assumed that there is some $\delta_0 >0$ such that 
$\sup_{\omega >\delta_0} |v_n^* (\omega ) v_{n'} (\omega )| <\infty$ and 
$\sup_{\omega >\delta_0} |d[v_n^* (\omega ) v_{n'} (\omega )]/d\omega |<\infty$ 
for all $n$ and $n'$. 
Then, if $\lambda$ is sufficiently small but not zero and 
the condition that $v_n(\omega_n)\neq 0$ for all $n\geq N-N_+ +1$ 
is satisfied, $H$ has no positive eigenvalues. 
\end{thm}

{\sl Proof} :  Under the assumption that $E>0$, 
we first consider the eigenvalue problem
\begin{equation}
\sum_{j=1}^{N} K_{ij} (E) c_{n j}
=\kappa_n (E, \lambda) c_{n i} ,
\label{eqn:3.2.70}
\end{equation}
for $i=1, 2, \ldots, N$, 
where 
$\{ c_{n i} \}_{i=1}^N$ is the normalized eigenvector 
corresponding the $n$-th eigenvalue $\kappa_n (E, \lambda)$ of $K(E)$. 
Then, by Theorem 4.3.1 in Ref. \cite{MatrixAnalysis}, one sees that
\begin{eqnarray}
\hspace{-5mm}
|\kappa_n (E,\lambda) -\omega_n | 
&\leq& \lambda^2 \max\{|\delta_1 (E) |,  |\delta_N (E) | \} 
\nonumber \\
&=& \lambda^2 \| D(E) \| \leq \lambda^2 \sup_{E>0} \| D(E) \| ,
\label{eqn:3.2.80}
\end{eqnarray}
for all $n$, where $\delta_n (E)$ is the $n$-th eigenvalue of $D(E)$. 
Note that from the assumption in the theorem and the propositions in the appendix, 
it holds that $\sup_{E>0} \| D(E) \| <\infty $. 
Therefore, by choosing $\lambda$ so that 
$|\lambda | < \lambda_a $, 
we have the fact that $|\kappa_n (E, \lambda)-\omega_n |<R_a$ for all $E>0$ and all $n$, 
and in particular $\kappa_n (E, \lambda) >0$ for all $E>0$ for all $n\geq N-N_+ +1$, 
where 
$\lambda_a = ( R_a /\sup_{E>0} \| D(E) \| )^{1/2}$ and 
\begin{equation}
R_a = 
\min \left\{ 
\omega_{N-N_+ +1}/3 ,
\min_{n, m} \{|\omega_n -\omega_m |/3 ~|~ n \neq m \} 
\right\} .
\label{eqn:3.2.85}
\end{equation}
The latter means that $\kappa_n (E,\lambda)$ only for $n\geq N-N_+ +1$ 
becomes a candidate for positive eigenvalue of $H$. 
Note that $\kappa_{N-N_+ }$ cannot be such a candidate even if $\omega_{N-N_+} =0$. 
Because in such a case, 
putting $\lambda_b = ( R_b /\sup_{E>0} \| D(E) \| )^{1/2}$, 
we find from Eq. (\ref{eqn:3.2.80}) that for $|\lambda |<\lambda_b $, 
$|\kappa_{N-N_+} (E,\lambda) | < R_b$ for all $E>0$. 
We here choose such a $R_b$ as to satisfy that $D(E') \geq 0 $ for all positive $E'< R_b$. 
Existence of such an $R_b$ is ensured by Eq. (\ref{eqn:a20}) 
in Proposition \ref{pp:a.1}. 
Then, from Theorem 4.3.1 in Ref. \cite{MatrixAnalysis} again, we have 
the estimation that 
\begin{equation}
-\lambda^2  \delta_N (E) 
\leq \kappa_{N-N_+} (E,\lambda) \leq 
-\lambda^2  \delta_1 (E) \leq 0,
\label{eqn:3.2.90}
\end{equation}
for all $E< R_b$. Hence, we conclude that if $|\lambda| <\lambda_b$, it holds that 
$\kappa_{N-N_+} (E,\lambda)<E$ for all $E>0$ \cite{E=0in multilevel}.

However, we can show that if we choose $\lambda$ sufficiently small, 
any such a $\kappa_n (E,\lambda)$ and eigenvector $\sum_i c_{ni} \ket{i}$ cannot 
satisfy Eq. (\ref{eqn:3.2.10}), no matter how well they satisfy Eqs. (\ref{eqn:3.2.50}) 
and (\ref{eqn:3.2.53}). 
To this end, let us look at Eq. (\ref{eqn:3.2.10}), which is rewritten as
\begin{eqnarray}
&&\hspace*{-10mm}
\left| \sum_{i=1}^{N} c_{ni} v_i (\kappa_n ) \right|^2
=
\left\| P_n (E, \lambda)  \sum_{i=1}^{N} v_i^* (\kappa_n ) \ket{i} \right\|^2
\label{eqn:3.2.100a} \\
&&\hspace*{-10mm}
=
|v_n^* (\kappa_n ) |^2
\nonumber \\
&&\hspace*{-10mm}
~~~+
\sum_{i=1}^{N} \bra{i} v_i (\kappa_n ) 
\bigl[ P_n (E, \lambda) -\ketbra{n}{n} \bigr]
\sum_{i'=1}^{N} v_{i'}^* (\kappa_n ) \ket{i'} ,
\label{eqn:3.2.100b}
\end{eqnarray}
where $P_n (E, \lambda)$ denotes the projection operator associated with 
the $n$-th eigenvalue $\kappa_n (E,\lambda)$. 
One sees that 
the first term on the right-hand side of 
Eq. (\ref{eqn:3.2.100b}) behaves as 
\begin{equation}
\lim_{\lambda \to 0} |v_n^* (\kappa_n (E,\lambda)) |^2 = |v_n^* (\omega_n ) |^2, 
\label{eqn:3.2.120}
\end{equation}
for all $E>0$ uniformly, because of Eq. (\ref{eqn:3.2.80}). 
From the assumption of the theorem, $|v_n^* (\omega_n ) |^2$ does not vanish. 
For the second term on the right-hand side of Eq. (\ref{eqn:3.2.100b}), 
we can use the result on the perturbation of 
the projection operator \cite{Kato(1966)}, which leads to the fact that 
\begin{equation}
P_n (E, \lambda) =
\ketbra{n}{n} 
+
\sum_{j=1}^{\infty} \lambda^{2j} P_n^{(j)}, 
\label{eqn:3.2.140}
\end{equation}
with 
\begin{equation}
P_n^{(j)}:= -\frac{1}{2\pi i} \oint_{\Gamma_n} 
(K_0 -\zeta)^{-1} [D(E) (K_0 -\zeta)^{-1} ]^j d\zeta, 
\label{eqn:3.2.150}
\end{equation}
where $\Gamma_n$ is the closed positively-oriented circle around $\zeta=\omega_n$ with 
radius $\min_{m(\neq n)} \{| \omega_n - \omega_m | /3 \}$. 
Series (\ref{eqn:3.2.140}) is ensured to converge uniformly 
for all $\lambda $ such that $|\lambda |< \min \{ \lambda_a, \lambda_b \}$, because 
\begin{equation}
\sup_{\zeta \in \Gamma_n} |\lambda|^2 \|D(E) \| \| (K_0 -\zeta)^{-1} \| 
< \lambda_a^2 /\lambda_n^2 \leq 1 ,
\label{eqn:3.2.160}
\end{equation}
where 
\begin{equation}
\lambda_n = \biggl[ \min_{m(\neq n)} \{| \omega_n - \omega_m | /3 \} \bigg/ \sup_{E>0} \|D(E) \| 
\biggr]^{1/2}. 
\label{eqn:3.2.165}
\end{equation}
From the assumption of no degeneracy among $\{ \omega_n \}_{n=1}^N$ 
and the discussion after Eq. (\ref{eqn:3.2.80}), for such a $\lambda$, 
all $\Gamma_n$'s are disconnected from each other, and 
there should be only one eigenvalue of $K$ in each circle. 
This leads to $\dim [P_n (E, \lambda) \mathbb{C}^N ] 
= \dim [\ketbra{n}{n} \mathbb{C}^N ] =1$, 
so that $\lambda =0$ is not an exceptional point \cite{Kato(1966)}. 
It is worth noting that $\lambda_n$ does not depend on $E$. 
Thus, the second term on the right-hand side of Eq. (\ref{eqn:3.2.100b}) 
is estimated as 
\begin{eqnarray}
&&\hspace{-5mm}
\sum_{i=1}^{N} \bra{i} v_i (\kappa_n ) 
\bigl[ P_n (E, \lambda) -\ketbra{n}{n} \bigr]
\sum_{i'=1}^{N} v_{i'}^* (\kappa_n ) \ket{i'}
\nonumber \\
&&\leq
\left\|
\sum_{i=1}^{N} v_i^* (\kappa_n ) \ket{i}
\right\|^2
\| P_n (E, \lambda) -\ketbra{n}{n} \|
\\
&&\leq
\left[
\sum_{i=1}^{N} \sup_{|\omega -\omega_n | <R_a}|v_i (\omega)|^2
\right]
\frac{(\lambda/ \lambda_n )^2 }{1-(\lambda/\lambda_n )^2 } 
\to 0, 
\label{eqn:3.2.110c}
\end{eqnarray}
as $\lambda \to 0$, for all $E>0$ uniformly, where it was used that 
$\sup_{\zeta \in \Gamma_n} \oint_{\Gamma_n} 
\| (K_0 -\zeta)^{-1} \| |d\zeta| \leq 2\pi$. 
Eq. (\ref{eqn:3.2.100b}) 
with the results (\ref{eqn:3.2.120}) and (\ref{eqn:3.2.110c}) 
means that Eq. (\ref{eqn:3.2.10}) is never satisfied 
for sufficiently small $\lambda$ with 
$|\lambda | < \min \{ \lambda_a, \lambda_b \}$, 
even if $\kappa_n (E,\lambda)=E$ holds. 
\qed

It is worth considering the opposite condition that $v_n (\omega_n )=0$. 
In this case, we could infer the existence of an eigenvalue inside the continuum, 
from the decay process arising from the pole $z_{{\rm p}, n}$. 
Indeed, if we recall the explicit form of the decay rate \cite{decayrate}, 
if the opposite condition holds, the decay rate comes small 
so that a much slower decay occurs. 
Then, one may associate such a behavior with the presence of a bound state 
\cite{example}, 
though it is not obvious whether this pole actually becomes an eigenenergy of $H$.

Let us now evaluate an explicit value of $\lambda$ for which 
there is no positive eigenvalue of $H$. 
Under the assumption of the analyticity of $v_n$, 
one sees that if $|\lambda | < \min \{ \lambda_a, \lambda_b \}$, 
Eq. (\ref{eqn:3.2.120}) is rewritten by using Eqs. (\ref{eqn:3.2.80}) and (\ref{eqn:3.2.165}) as
\begin{eqnarray}
&&\hspace{-5mm}\left| |v_n^* (\kappa_n (E,\lambda)) |^2 - |v_n^* (\omega_n ) |^2 \right|
\nonumber \\
&&\leq
\sup_{|\omega -\omega_n | <R_a} \left| \frac{d |v_n (\omega)|^2 }{d\omega} \right| 
|\kappa_n (E,\lambda) - \omega_n |
\label{eqn:3.2.200a}\\
&&=
\frac{\lambda^2}{ \lambda_n^2}
\min_{m (\neq n)} \{ |\omega_n-\omega_m|/3 \} 
\sup_{\omega >0} \left| \frac{d |v_n (\omega)|^2 }{d\omega} \right| .
\label{eqn:3.2.200b}
\end{eqnarray}
Therefore, by setting Eqs. (\ref{eqn:3.2.110c}) and (\ref{eqn:3.2.200b}) 
into Eq. (\ref{eqn:3.2.100a}), 
the left-hand side of Eq. (\ref{eqn:3.2.100a}) is ensured to be positive, 
and no positive eigenenergy of $H$ exists, 
providing that $\lambda$ is chosen to satisfy the $N_+ +1$ inequalities, 
\begin{equation}
|\lambda | < \min \{ \lambda_a, \lambda_b \}, 
\label{eqn:3.2.205}
\end{equation}
and 
\begin{eqnarray}
\hspace{-5mm}|v_n^* (\omega_n ) |^2 
&>&
\frac{\lambda^2}{\lambda_n^2} 
\min_{m (\neq n)} \{ |\omega_n-\omega_m|/3 \} 
\sup_{\omega >0} \left| \frac{d |v_n (\omega)|^2 }{d\omega} \right| 
\nonumber \\
&&+
\left[ \sum_{i=1}^{N} \sup_{|\omega -\omega_n | <R_a} |v_i (\omega)|^2 \right]
\frac{\lambda^2/\lambda_n^2 
 }{1-\lambda^2/\lambda_n^2  } ,
\label{eqn:3.2.210}
\end{eqnarray}
for $n = N-N_+ +1, \ldots, N$. 
By solving Eq. (\ref{eqn:3.2.210}) for $\lambda$ explicitly, 
Eqs. (\ref{eqn:3.2.205}) and (\ref{eqn:3.2.210}) are reduced into the single inequality 
\begin{equation}
|\lambda | < \min \{ \lambda_a, \lambda_b, 
\bar{\lambda}_{N-N_+ +1}, \ldots, \bar{\lambda}_N \}, 
\label{eqn:3.2.220}
\end{equation}
with
\begin{eqnarray}
&&\hspace*{-3mm}
\bar{\lambda}_n \hspace*{-1mm}
= \hspace*{-1mm}
\sqrt{
\frac{\lambda_n^2 }{2\beta_n}[ \alpha_n +\beta_n+\gamma_n - \sqrt{(\alpha_n +\beta_n+\gamma_n )^2-4\alpha_n \beta_n}] 
}
\nonumber \\
&& \hspace*{1mm} < \lambda_n ,
\label{eqn:3.2.230}
\end{eqnarray}
where $\alpha_n =|v_n^* (\omega_n ) |^2 $, 
$\beta_n
=\min_{m (\neq n)} \{ |\omega_n-\omega_m|/3 \} 
\sup_{\omega >0} \left| d |v_n (\omega)|^2 /d\omega \right| 
$, and 
$\gamma_n =\sum_{i=1}^{N} \sup_{|\omega -\omega_n | <R_a} |v_i (\omega)|^2$. 

In order to demonstrate Theorem \ref{thm:4.2}, 
we apply it to the spontaneous emission process for the hydrogen atom 
interacting with the electromagnetic field \cite{Facchi(1998)}. 
We suppose that 
$|n\rangle$ is the product state between 
the $(n+1)p$-state of the atom 
and the vacuum state of the field, 
and also $|\omega \rangle$ the product state between 
the $1s$-state of the atom and the one-photon state. 
Then, an initially excited atom is expected to make a transition to the ground state 
by emitting a photon. 
We treat the atom as a four-level system composed of the ground state and 
the three excited states: the $2p$, $3p$, and $4p$ state. 
The form factors corresponding to 
the $2p-1s$, $3p-1s$, and $4p-1s$ transitions were obtained as follows
\cite{Seke(1994),Facchi(1998),Miyamoto(2005)}, 
\begin{eqnarray}
\hspace*{-5mm}
v_1^* (\omega)
&=&
i \Lambda_1^{1/2} 
\frac{(\omega/\Lambda_1 )^{1/2}}{[1+(\omega/\Lambda_1 )^2]^2} ,
\label{eqn:3.2.320a} \\
v_2^* (\omega)
&=&
i 81 \Lambda_1^{1/2} 
\frac{(\omega/\Lambda_2 )^{1/2} [1+2(\omega/\Lambda_2 )^2]}
{128\sqrt{2} [1+(\omega/\Lambda_2 )^2]^3} ,
\label{eqn:3.2.320b} \\
v_3^* (\omega)
&=&
i 54 \sqrt{3} \Lambda_1^{1/2}  (\omega/\Lambda_3 )^{1/2}
\nonumber \\
&&
\times
\frac{45+146(\omega/\Lambda_3 )^2 +125(\omega/\Lambda_3 )^4}
{15625 [1+(\omega/\Lambda_3 )^2]^4} ,
\label{eqn:3.2.320c}
\end{eqnarray}
where $\Lambda_1=8.498 \times 10^{18} \ s^{-1}$, 
$\Lambda_2=(8/9)\Lambda_1 \ s^{-1}$, and 
$\Lambda_3=(10/12)\Lambda_1 \ s^{-1}$ are the cut off constants. 
One sees that these form factors satisfy all conditions required in Theorem \ref{thm:4.2}. 
The coupling constant is also given by $\lambda^2 =6.435 \times 10^{-9}$. 
The eigenvalues of $H_0$ are given by 
$\omega_n = \frac{4}{3}\Omega [1-(n+1)^{-2}]$ with 
$\Omega = 1.55 \times 10^{16} s^{-1}$, all of which are embedded in the energy continuum. 
The Hamiltonian (\ref{eqn:2.60}) is then derived under 
the four-level approximation (i.e., $N=N_+=3$) and the rotating-wave approximation. 
The various parameters are numerically obtained as follows: 
$R_a=| \omega_2 - \omega_3 | /3 =(7/324)\Omega$, 
$\sup_{E>0} \|D(E) \|=-\delta_1 (E) =11.332 \Lambda_1$ at $E=0.6145 \Lambda_1$, 
$\lambda_1^2=5.45 \times 10^{-3}\Omega/\Lambda_1$, 
$\lambda_2^2=\lambda_3^2=\lambda_a^2=1.91 \times 10^{-3}\Omega/\Lambda_1$, 
$\alpha_1=1.82 \times 10^{-3}\Lambda_1$, 
$\alpha_2=4.87 \times 10^{-4}\Lambda_1$, 
$\alpha_3=1.99 \times 10^{-4}\Lambda_1$, 
$\beta_1=6.17 \times 10^{-2}\Omega$, 
$\beta_2=4.87 \times 10^{-3}\Omega$, 
$\beta_3=1.88 \times 10^{-3}\Omega$, 
$\gamma_1=2.45 \times 10^{-3}\Lambda_1$, 
$\gamma_2=3.04 \times 10^{-3}\Lambda_1$, 
$\gamma_3=2.45 \times 10^{-3}\Lambda_1$, 
from which Eq. (\ref{eqn:3.2.230}) reads 
$\bar{\lambda}_1^2=4.18 \times 10^{-6}$, 
$\bar{\lambda}_2^2=5.01 \times 10^{-7}$, and 
$\bar{\lambda}_3^2=2.14 \times 10^{-7}$. 
Then, it follows that 
\begin{equation}
\min \{ \lambda_a^2, \bar{\lambda}_{1}^2, \bar{\lambda}_{2}^2, \bar{\lambda}_3^2 \}
=
\bar{\lambda}_3^2 
>\lambda^2 ,
\label{eqn:3.2.240}
\end{equation}
and thus Eq. (\ref{eqn:3.2.220}) holds. 
This conclusion indicates that 
the intrinsic values of the parameters characterizing the system 
does not allow any bound state.
In fact, we have not observed any such state. 
It is worth noticing that the upper bound estimated in Eq. (\ref{eqn:3.2.240}) 
is dominated by the factor $\lambda_3^2$, roughly speaking, 
the minimum level-spacing over the maximum cut off constant.


\section{Concluding remarks} 
\label{sec:5}

We have considered the eigenvalue problem for unstable multilevel systems, 
on the basis of the $N$-level Friedrichs model, 
where the eigenenergies are supposed outside or possibly inside the continuum. 
The outside case is essentially determined by 
the location of the discrete level $\omega_n$ of the free Hamiltonian and 
the strength of the coupling constant $\lambda$. 
If $\omega_n$ lies outside the continuum, 
the corresponding eigenvalue always lies below $\omega_n$. 
If $\omega_n$ lies inside the continuum, by choosing a $\lambda$ large enough 
the eigenvalue originating from $\omega_n$ can emerge from the continuum. 
Such behaviors are similar to those seen in single-level cases, 
however, this is not the case if the form factors $v_n$ are linearly dependent. 
On the other hand, we have shown the absence of the eigenvalue lying inside the continuum 
in the weak coupling cases, 
under the condition that $v_n (\omega_n )\neq 0$ if $\omega_n $ lies inside the continuum. 
This statement is just an extension of Lemma 2.1 in Ref. \cite{Davies(1974)}, 
where only identical form factors 
were considered, 
and the upper bound for $|\lambda |$ required in the lemma was not estimated. 
We have evaluated this upper bound in our case, which proves to be proportional to 
the minimum level-spacing over the maximum cut off constant. 
Hence, comparing this value with the actual $\lambda$, 
one can check at least the absence of the eigenvalue,  
even in the case that one cannot evaluate the reduced resolvent explicitly. 
At first sight, the normalization condition, i.e., Eq. (\ref{eqn:3.2.10}), seems not 
necessarily to require the zeros of the form factors 
for a presence of a bound-state eigenenergy inside the continuum, 
though it is misplaced in weak-coupling regimes. 
However, we still do not have a definite answer to this matter in other coupling regimes 
where the multilevel effect may allow a presence of a bound-state eigenenergy inside the continuum 
without zeros of the form factors.


\section*{Acknowledgments}

The author would like to thank Professor I.\ Ohba 
and Professor H.\ Nakazato for useful comments. 
He would also like to thank the Yukawa Institute for Theoretical Physics 
at Kyoto University, where this work was initiated during 
the YITP-04-15, 
Fundamental Problems and Applications of Quantum Field Theory. 
This work is partly supported by a Grant for the 21st Century COE Program 
at Waseda University from the Ministry of Education, Culture, 
Sports, Science and Technology, Japan. 


\appendix*
\section{}

In this section, we present the two Propositions \ref{pp:a.1} and \ref{pp:a.4}. 
The former and the latter state that the behavior of the energy shift $D(E)$ 
at small and large energies is quite regular without any divergence, respectively, 
under some form-factor conditions that are often satisfied by actual systems.

\begin{pp}\label{pp:a.1}
Suppose that the function $\eta (\omega)$ belonging to $L^1 ([0, \infty ))$ is of the form 
\begin{equation}
\eta (\omega ) := \omega^p r(\omega ), 
\label{eqn:a10}
\end{equation}
where $p>0$ and $r(\omega)$ is a $C^1$-function defined in $[0,\infty)$. 
It then holds that $\eta (\omega)/\omega \in L^1 ([0, \infty ))$ and 
\begin{equation}
\int_{0}^{\infty} \frac{\eta (\omega)}{\omega} d\omega
=
\lim_{E \uparrow 0} \int_{0}^{\infty} \frac{\eta (\omega)}{\omega -E} d\omega
=\lim_{E \downarrow 0} P\int_{0}^{\infty} \frac{\eta (\omega)}{\omega -E} d\omega .
\label{eqn:a20}
\end{equation}
\end{pp}

{\sl Proof} : From the proof of Proposition 3.2.2 in Ref. \cite{Exner(1985)}, 
the principal value of the integral on the right-hand side is written by the absolutely 
integrable function as follows 
\begin{equation}
P\int_{0}^{\infty} \frac{\eta (\omega)}{\omega -E} d\omega 
=\int_{0}^{\infty} \frac{\eta (\omega)-\eta (E)\varphi_{\delta}(\omega-E)}{\omega -E} d\omega ,
\label{eqn:a30}
\end{equation}
for all $E>0$, where $\varphi_{\delta}(\omega)$ is a $C_0^{\infty}$-function 
with support $[-\delta , \delta ]$ ($0< \forall \delta < E$), even with respect to the origin, 
and such that $\varphi_{\delta}(0)=1$. 
In the following, we choose $\varphi_{\delta}(\omega)=\exp[1-1/(1-(\omega/\delta)^2)]$ 
for $\omega \in (-\delta , \delta )$ or $0$ otherwise, and $\delta=E/2$. 
On the other hand, since from the assumption (\ref{eqn:a10}) $\eta (\omega) /\omega$ is absolutely integrable, the first equality in Eq. (\ref{eqn:a20}) is obvious. 
Therefore, it is sufficient to show that 
\begin{equation}
\lim_{E \downarrow 0} 
\int_{0}^{\infty} 
\left[ 
\frac{\eta (\omega)}{\omega}
-\frac{\eta (\omega)-\eta (E)\varphi_{\delta}(\omega-E)}{\omega -E} 
\right] 
d\omega =0.
\label{eqn:a40}
\end{equation}
Note that the above integrand can be rewritten as 
\begin{eqnarray}
&&\hspace*{-10mm}\frac{\eta (\omega)}{\omega}
-\frac{\eta (\omega)-\eta (E)\varphi_{\delta}(\omega-E)}{\omega -E} 
\nonumber \\
&&\hspace*{-10mm}=
-E\frac{\eta (\omega)}{\omega(\omega-E)}
+\frac{\eta (E)\varphi_{\delta}(\omega-E)}{\omega -E} 
\label{eqn:a50} \\
&&\hspace*{-10mm}=
\frac{\eta (E)\varphi_{\delta}(\omega-E)}{\omega}
-E\frac{\eta (\omega)-\eta (E)\varphi_{\delta}(\omega-E)}{\omega(\omega -E)} .
\label{eqn:a60} 
\end{eqnarray}
Let us first consider the case where $\omega \in I:= (0, E/2] \cup [3E/2, \infty)$. 
Then, since $\varphi_{\delta} (\omega-E)=0$, we can use Eq. (\ref{eqn:a50}) to 
estimate the integrand: 
\begin{equation}
\left|
E\frac{\eta (\omega)}{\omega(\omega-E)} 
\right|
\leq 
2 \left|
\frac{\eta (\omega)}{\omega} 
\right| , 
\label{eqn:a70}
\end{equation}
where the right-hand side is absolutely integrable and independent of $E$. 
Furthermore, it follows that 
$\lim_{E \downarrow 0} E \chi_{I} (\omega ) \eta (\omega)/[\omega(\omega-E)] =0$ for every $\omega \in (0, \infty )$, where $ \chi_{I} (\omega ) = 1$ ($\omega \in I$) or $0$ ($\omega \notin I$), being the characteristic function. 
Thus, by the dominated convergence theorem, we can see that 
\begin{equation}
\lim_{E \downarrow 0} 
\left(
\int_{0}^{E/2} +
\int_{3E/2}^{\infty} 
\right)
E\frac{\eta (\omega)}{\omega(\omega-E)}
d\omega 
= 0.
\label{eqn:a80}
\end{equation}
For $\omega \in (E/2, 3E/2 )$, we can use Eq. (\ref{eqn:a60}). 
The integration of the first term of Eq. (\ref{eqn:a60}) is estimated  by 
\begin{equation}
\begin{array}{l}
{\displaystyle
\left|
\int_{E/2}^{3E/2}
\frac{\eta (E)\varphi_{\delta}(\omega-E)}{\omega}
d\omega 
\right|}
\\
{\displaystyle
\leq
\frac{\eta (E)}{E/2}
\int_{E/2}^{3E/2}
\varphi_{\delta}(\omega-E)
d\omega 
=
\eta (E)
\int_{-1}^{1}
\varphi_{1}(x)
dx \to 0, 
}
\end{array}
\label{eqn:a85}
\end{equation}
as $E \downarrow 0$. 
The second term of Eq. (\ref{eqn:a60}) is also estimated by 
\begin{eqnarray}
&&
|\eta(\omega)-\eta(E)\varphi_{\delta}(\omega-E)| 
\nonumber \\
&& \leq 
|\eta(\omega) -\eta(E)|+|\eta(E)||1-\varphi_{\delta}(\omega-E)| .
\label{eqn:a90}
\end{eqnarray}
The integration of the first term on the right-hand side right-hand side of the above is evaluated as 
\begin{eqnarray}
&&
\int_{E/2}^{3E/2} 
E\frac{|\eta (\omega)-\eta (E)|}{\omega|\omega -E|} d\omega
\nonumber \\
&&\leq
(\ln 3) E \sup_{E/2 \leq \omega \leq 3E/2} \left| \eta^{\prime}(\omega) \right|
\label{eqn:a100} \\
&&\leq
(\ln 3) E 
\left[
p E^{p-1}
\max\{ ({\textstyle \frac{1}{2}})^{p-1}, ({\textstyle \frac{3}{2}})^{p-1}\}
\sup_{\omega \in [0, 3E/2]} \left| r(\omega) \right|
\right.
\nonumber \\
&&~~~
+
\left.
\left(\frac{3E}{2}\right)^{p}
\sup_{\omega \in [0, 3E/2]} \left| r^{\prime}(\omega) \right|
\right] 
\to 0 \mbox{ as } E \downarrow 0,
\label{eqn:a120} 
\end{eqnarray}
where the prime on $\eta^{\prime} (\omega)$ 
implies the differentiation of $\eta (\omega)$ and so on. 
The integral corresponding to the last term on the right-hand side 
of Eq. (\ref{eqn:a90}) is also estimated as 
\begin{eqnarray}
&&\hspace*{-5mm}
\int_{E/2}^{3E/2} 
E\frac{|\eta (E)||1-\varphi_{\delta}(\omega-E)|}{\omega|\omega -E|} d\omega
\nonumber \\
&&\hspace*{-5mm}\leq
(\ln 3) E |\eta (E)| \sup_{E/2 \leq \omega \leq 3E/2} \left| \varphi_{\delta}^{\prime} (\omega-E) \right|
\label{eqn:a130} \\
&&\hspace*{-5mm}=
2 (\ln 3) |\eta (E)| \sup_{|x| \leq 1} \left| \varphi_{1}^{\prime}(x) \right|
\rightarrow 0 ~~~(E \downarrow 0). 
\label{eqn:a140} 
\end{eqnarray}
Thus, we can obtain 
\begin{equation}
\lim_{E\downarrow 0} 
\int_{E/2}^{3E/2} 
E\frac{\eta (\omega)-\eta (E)\varphi_{\delta}(\omega-E)}{\omega(\omega -E)} 
d\omega=0. 
\label{eqn:a150} 
\end{equation}
Equations (\ref{eqn:a80}), (\ref{eqn:a85}), and (\ref{eqn:a150}) 
mean the completion of the proof of (\ref{eqn:a40}). \qed

\begin{pp}\label{pp:a.4}
Suppose that the function $\eta (\omega)$ belongs to 
$L^1 ([0, \infty )) \cap C^1 ([0, \infty ))$, and satisfies 
that $\sup_{\omega \geq \delta_0} |\eta (\omega)| <\infty$ 
and $\sup_{\omega \geq \delta_0} |\eta' (\omega)|<\infty $ for some $\delta_0 >0$. 
Then, 
\begin{equation}
\sup_{E>\delta_0} 
\left|
P\int_0^\infty \frac{\eta (\omega)}{\omega -E} d\omega
\right|
<\infty .
\label{eqn:a290}
\end{equation}
\end{pp}

{\sl Proof} : 
To examine this integral, we use the expression (\ref{eqn:a30}) 
and divide the interval $[0, \infty )$ into 
$I_{\delta, E} = [E-\delta , E+\delta ] $ and 
$\overline{I_{\delta, E}} = [0, \infty )\setminus I_{\delta, E} $, again, 
where we assume $\delta_0 >\delta>0$. 
In the latter interval, it is estimated that 
$\chi_{\overline{I_{\delta, E}}} (\omega) 
|\eta (\omega)/(\omega -E)| \leq |\eta (\omega)|/\delta \in L^1 ([0, \infty ))$. 
Then, 
\begin{equation}
\sup_{E>\delta_0} 
\left|
\int_0^\infty 
\chi_{\overline{I_{\delta, E}}} (\omega) 
\frac{\eta (\omega)}{\omega -E} d\omega
\right|
\leq
\frac{1}{\delta} \int_0^\infty |\eta (\omega)| d\omega <\infty .
\label{eqn:a300}
\end{equation}
In the former interval, the integrand in Eq. (\ref{eqn:a30}) is evaluated as 
\begin{eqnarray}
\left|
\frac{\eta (\omega)-\eta (E) \varphi_\delta (\omega-E)}{\omega -E}
\right|
&\leq&
\sup_{\omega \in I_{\delta, E} } |\eta' (\omega)| 
\nonumber \\
&&
+|\eta (E)| 
\sup_{|\omega |\leq \delta } |{\varphi_\delta}' (\omega)| ,
\label{eqn:a310}
\end{eqnarray}
which results in  
\begin{eqnarray}
&&
\sup_{E>\delta_0} 
\left|
\int_0^\infty \chi_{I_{\delta, E}} (\omega) 
\frac{\eta (\omega)-\eta (E) \varphi_\delta (\omega-E)}{\omega -E}
\right|
\nonumber \\
&&\leq
2\delta 
\left[
\sup_{E>\delta_0} |\eta' (E)| +\sup_{E>\delta_0} |\eta (E)| 
\sup_{|\omega |\leq \delta } |{\varphi_\delta}' (\omega)|
\right] <\infty,
\nonumber \\
&&
\label{eqn:a320}
\end{eqnarray}
where we used the assumption for $\eta (\omega)$ in the statement. 
Incorporating Eq. (\ref{eqn:a300}) 
with Eq. (\ref{eqn:a320}), Eq. (\ref{eqn:a290}) is obtained. 
\qed



\begin{thebibliography}{99}
\bibitem{pointspectrum}
The eigenvalue problem mentioned here is mathematically equivalent 
to finding the point spectrum of 
$H$ 
defined on the Hilbert space ${\cal H}=\mathbb{C}^N \oplus L^2 ([0, \infty))$.
\bibitem{c_1}
Note that $c_1 \neq 0$. Indeed, if $c_1=0$, 
Eq. (\ref{eqn:3.20}) reads $(\omega-E)f(\omega)=0$. However, since we require that 
$f(\omega) \in L^2 (0, \infty )$, not the distribution, 
this relation results in a trivial solution, $f(\omega)=0$.
\bibitem{inequality}
Theorem 4.3.1 concerns the perturbation of the eigenvalue of a hermitian matrix: 
Suppose that $A$ and $B$ are $N\times N$ hermitian matrices. Then it holds that 
$\lambda_k (A) + \lambda_1 (B) \leq \lambda_k (A+B) \leq \lambda_k (A) + \lambda_N (B)$,  
where $\lambda_k (A)$ denotes the $k$-th eigenvalue of $A$, and so forth. 
Note that $\lambda_1 (B)$ and $\lambda_N (B)$ are just the minimum and maximum eigenvalues of $B$, 
respectively.
\bibitem{zeros}
Note that when $v_1 (\omega)$ is differentiable, 
the condition (\ref{eqn:3.2.55a}) means 
that $v_1 (\omega)$ comes to $0$ smoothly around $\omega=E$. 
Such zeros of $v_1 (\omega)$ could be yielded instead by 
the ``point-like'' gaps of $\rho (\omega)$. 
If $v_1 (\omega)$ behaves like $\lim_{\omega \to E} v_1 (\omega)\neq 0$ but $v_1 (E)=0$ 
discontinuously, Eq. (\ref{eqn:3.30}) never holds.
\bibitem{E=0}
For $E=0$, one can replace $P\int_{0}^{\infty} \frac{|v_1 (\omega)|^2}{\omega} d\omega$ 
by $\int_{0}^{\infty} \frac{|v_1 (\omega)|^2}{\omega} d\omega$ under the condition 
$v_1 (0)=0$.
\bibitem{E=0in multilevel}
This argument does not exclude the possibility of 
$\kappa_{N-N_+} (0, \lambda)$ being the zero-energy eigenvalue of $H$ 
for $|\lambda | <\min \{ \lambda_a ,\lambda_b \}$. 
In this case, Eq. (\ref{eqn:3.2.90}) implies that $\delta_1 (0)=\sigma_1 (0)=0$, 
so that the linear dependency among $v_n (\omega)$ is at least required. 
In addition, the discussion in Sec. \ref{sec:3} tells us that 
$\kappa_{N-N_+} (E)=0$ for all $E<0$ should hold.
\bibitem{decayrate}
In weak-coupling cases, the pole $z_{{\rm p}, n}$ originating from $\omega_n$ is of the form 
$z_{{\rm p}, n} = \omega_n -\lambda^2 D_{nn}(\omega_n ) -i\pi \lambda^2 |v_n (\omega_n )|^2 
+O(\lambda^3)$ \cite{Exner(1985),Antoniou(2003)}. 
The decay rate is then given by $-2 {\rm Im} z_{{\rm p}, n}$.
\bibitem{example}
For a system with identical form factors, which have a common zero, 
we can find an appropriate $\lambda$ to make such a zero an actual eigenvalue.
\end{thebibliography}
\end{document}